\documentstyle[12pt,epsf]{article} 
\input epsf

\newcommand{\frat}[2]{\frac{\textstyle #1}{\textstyle #2}}
\newcommand{\dmn}[2]{\mbox{$#1\!\cdot\! 10^{#2}\,$}}
\newcommand{\grpicture}[1]
{\epsfxsize=200pt 
    \hspace{5cm} \parbox{200pt}{\epsfbox{#1.ps}} \vspace{5mm}
}

\begin{document}
\begin{center}
{\Large 
\bf Quark induced excitations of the instanton liquid}\\
\vspace{0.5cm}
{S.V. Molodtsov, A.M. Snigirev$^\dagger$, G.M. Zinovjev$^\ddagger$}\\
\vspace{0.5cm} 
 {\small\it State Research Center,
Institute of Theoretical and Experimental Physics,
117259, Moscow, Russia\\ 
$^\dagger$Skobeltsyn Institute of Nuclear Physics, Moscow State
University,
119899, Moscow, Russia\\
$^\ddagger$Fakult\"at f\"ur Physik, Universit\"at Bielefeld,
D-33501, Bielefeld, Germany\\
$^\ddagger$Bogolyubov Institute for Theoretical Physics,\\
National Academy of Sciences of Ukraine, 
UA-03143, Kiev, Ukraine} 
\end{center}
\vspace{0.5cm}
\begin{abstract}
The selfconsistent approach to the quark interactions in the instanton liquid 
is developed within the tadpole approximation  calculating the basic functional
integral. The effective Lagrangian obtained includes colourless scalar field
interacting with quarks. The origin of this dynamical field as an interaction 
carrier in soft momentum region is discussed.
\end{abstract}
\vspace{0.5cm}
PACS: 11.15 Kc,  12.38-Aw
\\
\\ 
\section*{I. Introduction}
Nowadays, there are no doubts the model of QCD vacuum as the instanton liquid 
(IL) is the most practical instrument on the chiral scale of QCD. It provides, 
as the lattice calculations recently confirmed, not only the theoretical 
background for describing spontaneous chiral symmetry breaking (SCSB) but
is mostly powerful in the phenomenology of the QCD vacuum and in the physics
of light quarks while considered to propagate by zero modes arising
from instantons. The origin of gluon and chiral condensates turns out in
this picture easily understandable and both are quantitatively calculated
getting very realistic values defined by $\Lambda_{QCD}$ and parameters of
instanton and anti-instanton ensemble, for example, 
$-i~\langle\psi^\dagger\psi\rangle\sim -(250~MeV)^3$. Moreover, the scale
for dynamical quark masses, $M\sim 350~MeV$, naturally appears and pion
decay constant, $F_\pi\sim 100~MeV$, is then transparently calculated.

Another significant advantage of this approach is that the initial formulation
starts basically from the first principles and subsequent approximations
being well grounded and reliably controlled are plugged in \cite{1},
\cite{2}. It becomes clear especially in recent years when the impressive
progress has been reached in understanding the instanton physics on the lattice
\cite{Negele}. Further we are summarizing several things we have learned 
thinking of the IL theory and trying to answer the challenging questions.

Let us start on that stage of the IL approach when its generating functional 
has already been factorized as 
$$ 
{\cal Z}~=~{\cal Z}_g~\cdot~{\cal Z}_\psi~,
$$
where eventually the factor ${\cal Z}_{g}$ provides nontrivial gluon 
condensate while the fermion part ${\cal Z}_{\psi}$ is responsible to 
describe the chiral
condensate in instanton medium and its excitations. It is usually
supposed the functional integral of ${\cal Z}_{g}$ is saturated by 
the superposition of the pseudo-particle (PP) fields which are the
Euclidean solutions of the Yang-Mills equations called the (anti-)instantons
\begin{equation}
\label{2} 
A_\mu(x)=\sum_{i=1}^N A_\mu(x;\gamma_i)~.
\end{equation}
Here $A_\mu(x;\gamma_i)$ denotes the field of a single (anti-)instanton
in singular gauge with $4N_c$ (for the $SU(N_c)$ group) coordinates 
$\gamma=(\rho,~z,~U)$ of size $\rho$ centred at the coordinate $z$
and colour orientation defined by the matrix $U$. The nontrivial bloc of
corresponding $N_{c} \times N_{c}$ matrices of PP is a part of potential
\begin{equation}
\label{3}
A_\mu(x;\gamma)=\frat{\bar \eta_{a\mu\nu}}{g}
\frat{y_\nu}{y^2}\frat{\rho^2}{y^2+\rho^2}~U^\dagger \tau_a~U,~~y=x-z~,
~a=1,2,3~,
\end{equation}
where $\tau_a$ are the Pauli matrices, $\eta$ is the 't Hooft
symbol \cite{3}, $g$ is the coupling constant
and for anti-instanton $\bar \eta \to \eta$. For the sake
of simplicity we do not introduce the distinct symbols for instanton
($N_{+}$) and anti-instanton ($N_{-}$) and consider topologically neutral
IL with $N_+=N_-=N/2$. Formulating the variational principle
the practical estimate of ${\cal Z}_{g}$ was found \cite{2} 
$$
{\cal Z}_g~\simeq~e^{-\langle S\rangle}
$$
with the action of IL defined by the following additive functional
{\footnote{
In fact, the additive property results from the supposed homogeneity of
vacuum wave function in metric space. Eq. (\ref{s}) looks like a formula of
classical physics although it describes the ground state of quantum
instanton ensemble. Intuitively clear, this definition will be still
valid even when the wave function is nonhomogeneous with the
nonuniformity scale essentially exceeding average instanton size or, 
precisely speaking, being larger (or of the order) than average size of 
characteristic saturating field configuration. Then each instanton liquid 
element of such a distinctive size will provide a partial contribution 
depending on the current state of IL, see next Section.}}
\begin{equation}
\label{s}
\langle S\rangle=\int d z \int d\rho~n(\rho)~s(\rho)~.
\end{equation}
The integration should be performed over the IL volume $V$
along with averaging the action per one instanton  
\begin{equation}
\label{si}
s(\rho)=\beta(\rho)+5 \ln(\Lambda\rho)-\ln \widetilde \beta^{2N_c}
+\beta \xi^2\rho^2\int d\rho_1~n(\rho_1)\rho_1^{2},
\end{equation}
weighted with instanton size distribution function
\begin{equation}
\label{nrho}
n(\rho)=C~e^{-s(\rho)}=C~\rho^{-5} \widetilde\beta^{2 N_c} 
e^{-\beta(\rho)-\nu \rho^2/\overline{\rho^2}}~,
\end{equation}
$$
\nu=\frat{b-4}{2},~b=\frat{11~N_c-2~N_f}{3}~,
$$
where
$\overline{\rho^2}=\int d\rho~\rho^2~n(\rho)/n=
\left(\frat{\nu}{\beta \xi^2 n}\right)^{1/2},
~n=\int d\rho~n(\rho)=N/V$ 
and $N_f$ is the number of flavours.
The constant $C$ is defined by the variational maximum principle in the
selfconsistent way and $\beta(\rho)=\frat{8\pi^2}{g^2}=
-\ln C_{N_c}-b \ln(\Lambda \rho)
~(\Lambda=\Lambda_{\overline{MS}}=0.92 \Lambda_{P.V.})$ 
with constant $C_{N_c}$ depending 
on the renormalization scheme, in particular, here 
$C_{N_c}\approx\frat{4.66~\exp(-1.68 N_c)}
{\pi^2 (N_c-1)!(N_c-2)!}$. The parameters $\beta=\beta(\bar\rho)$ and 
$\widetilde \beta=\beta+\ln C_{N_c}$ are 
fixed at the characteristic scale $\bar\rho$ (an average instanton size).
The constant $\xi^2=\frat{27}{4}\frat{N_c}{N_c^{2}-1} \pi^2$ 
characterizes, in a sense, the PP interaction and Eqs. (\ref{s}),(\ref{si}) 
and (\ref{nrho}) describe the equilibrium state of IL.
The minor modification of variational maximum principle (see 
Appendix) leads to the explicit form of the mean instanton size 
$\bar\rho\Lambda=\exp\left\{-\frat{2N_c}{2\nu-1}\right\}$ and, therefore, 
to the direct definition of the IL parameters unlike the  
conventional variational principle \cite{2} which allows one to extract
those parameters solving numerically the transcendental equation only.

The quark fields are considered to be {\it influenced} by the certain
stochastic ensemble of PPs, Eq. (\ref{2}), while calculating the quark 
determinant   
$$
{\cal Z}_\psi~\simeq~\int D\psi^\dagger D\psi~~\langle\langle~
e^{S(\psi,\psi^\dagger,A)}~\rangle\rangle_A~.
$$
Besides, dealing with the dilute IL (small characteristic  
packing fraction parameter $n~\bar\rho^4$) one neglects the correlations 
between PPs and utilizes the approximation of $N_c\to\infty$ where the planar 
diagrams only survive. In addition the fermion field action is 
approached by the zero modes which are the solutions of the Dirac 
equation $i(\hat D(A_\pm)+m)~\Phi_\pm=0$ in the field of (anti-)instanton 
$A_\pm$, i.e. 
\begin{equation}
\label{gf}
\left[\Phi_\pm(x)\right]_{ic}=\frat{\rho}{\sqrt{2}\pi |x|(x^2+\rho^2)^{3/2}}
\left[\hat x~\frat{1\pm\gamma_5}{2}\right]_{ij}\varepsilon_{jd}
~U_{dc}
\end{equation}
with the colour $c,d$ and the Lorentz $i,j$ indices and the antisymmetric 
tensor $\varepsilon$.  
In particular, at $N_f=1$ the quark determinant reads \cite{2}
\begin{equation}
\label{6}
{\cal Z}_\psi~\simeq~\int D\psi^\dagger D\psi ~ 
\exp\left\{\int d x~\psi^\dagger(x)~i\hat\partial\psi(x)\right\}~
\left(\frat{Y^+}{VR}\right)^{N_+}~
\left(\frat{Y^-}{VR}\right)^{N_-}~,
\end{equation}
$$Y^{\pm}=i\int dz~dU~d\rho~n(\rho)/n
\int dxdy~\psi^\dagger(x)~i\hat\partial_x \Phi_\pm(x-z)~
\Phi^\dagger_{\pm}(y-z)~i\hat\partial_y~\psi(y)~, 
$$
where the factor $R$ makes the result dimensionless and is also fixed by the
saddle point calculation. Much more accurate result for the Green function of
quark in the PP ensemble \cite{lee} allows one to calculate the generating
functional even beyond the chiral limit and the simple zero mode approximation
turns out amazingly fruitful again to develop (quantitatively) the  
low energy phenomenology of light quarks \cite{mu}.

Thus, the IL approach at the present stage of its development looks very 
indicative, well theoretically grounded and reasonably adjusted 
phenomenologically. The proper form of generating functional obtained
and its reasonable parameter dependence provide with enough predictive
power and justify, hence, the approximations made. Generally, it leads
to conclude the quark feedback upon the instanton background is pretty
limp and could be perturbatively incorporated as a small variation of
instanton liquid parameters $\delta n$ and $\delta \rho$ around their 
equilibrium values of $n$ and $\bar\rho$ being in full analogy with the 
description of chiral condensate excitations. Indeed, the result of 
nontrivial calculation of the functional integral (treating the zero quark 
modes in the fermion determinant substantially from physical view point) 
takes down to 'encoding' the IL state with just those two parameters. 
Moreover, the IL density appears in the approach via the packing fraction 
parameter $n\bar\rho^4$ only (clear from dimensional analysis) what means one 
independent parameter existing in practice. It is just the instanton size. The
analysis of the {\it quark and IL interaction} is addressed in this paper
developing our idea \cite{we} of phononlike excitations of IL resulting from 
the adiabatic changes of the instanton size. Hence, we describe further the
{\it quark feedback} dealing with these deformable (anti-)instanton 
configurations which are the field configurations Eq. (\ref{3}) characterized 
by the size $\rho$ depending on $x$ and $z$, i.e. $\rho\to\rho(x,z)$.

The paper is organized as follows: in Section II we discuss the
modification of quark determinant when the functional integral is
saturated by the deformable modes at one flavour. 
Then in Section III we develop the proper calculation (tadpole 
approximation) which is based on the saddle point method. 
Section IV is devoted to the 
generalization for the multiflavour picture. The paper includes also Appendix 
where the fault finding reader gets a chance to control the explicit formulae 
for the IL parameters and to improve our calculations if is able.
 
\section*{II. Dynamics of phononlike excitations}
Apparently, the essence of what we discuss here could be illuminated in the 
following way. Saddle point method calculation of the functional integral 
implies the treatment of the action extremals which are the solutions of 
classical field equations. For the case in hands the action 
$S[A,\psi^\dagger,\psi]$ is constructed including the gluon fields $A_{\mu}$, 
(anti-)quarks $\psi^\dagger,\psi$ and extremals which are given by the 
solutions of the consistent system of the Yang-Mills and Dirac equations. As 
a trial configuration in the IL theory the superposition of (anti-)instantons
which is the approximate solution of the Yang-Mills equations (with no
backreaction of the quark fields) and a background field for the Dirac
equation is simultaneously considered. We believe it is reasonable to make
use the deformable (crumpled) (anti-)instantons $A_{\pm}(x;\gamma(x))$
as the saturating configurations. They just admit of varying the
parameters $\gamma(x)$ of the Yang-Mills sector of the initial consistent 
system in order to describe the influence of quark fields in the appropriate 
variables for the quark determinant. 

Taking the action in the form $S[A_{\pm}(x,\gamma(x)),\psi^\dagger,\psi]$ we 
would receive the corresponding variational equation for the deformation field 
$\gamma(x)$ which approaches most optimally (as to the action extremum) PP
at nonzero quark fields. In the field theory, for example, the monopole
scattering \cite{man}, the Abrikosov vortices scattering \cite{serg} are
treated in a similar way. 
Actually, the unstationary picture seems more adequate as a specific role 
of instanton-anti-instanton annihilation channel prompts and an investigation 
of two-body colour problem teaches \cite{we1}. 

In practice, for IL we avoid the difficulties which 
come with solving the variational equations if we consider the long wavelength 
excitations only (with the wave length $\lambda$ much larger than the 
characteristic instanton size $\bar\rho$). Indeed, it can be done because
we are searching the kinetic energy of the deformation fields 
{\footnote{Then we are allowed to take the slowly changing deformation field
beyond the integral while calculating the action of deformed instanton.}} 
(the one particle contributions) and consider the pair interaction which 
develops the contact interaction form being calculated in the adiabatic regime
\cite{we}. 

Let us remember here that deriving Eq. (\ref{s}) we should average
over the instanton positions in a metric space. Clearly, the characteristic
size of the domain $L$ which has to be taken into account should exceed the 
mean instanton size $\bar\rho$. But at the same time it should
not be too large because the far ranged elements of IL are not 'causally'
related. The ensemble wave function is expected to be homogeneous (every
PP contributes to the functional integral being weighted with a factor
proportional to $\sim 1/V,~V=L^4$) on this scale. The characteristic
configuration which saturates the functional integral is taken as the
superposition Eq. (\ref{2}) with $N$ number of PP in the volume $V$. 
It is easy
to understand that because of an additivity of the functional, Eq.(\ref{s}) 
describes properly even non-equilibrium states of IL when the distribution
function $n(\rho)$ does not coincide with the vacuum one and, moreover, allows 
us to generalize this action for the non-homogeneous liquid 
when the size of the
homogeneity obeys the obvious requirement $\lambda\ge L>\bar\rho$. In
particular, the ensemble of deformable PPs as saturating the functional
integral obeys these constraints if the corrections coming with the
deformation fields $g_{\mu\nu}$ are small and the long wavelength
excitations (on the scale of the instanton size $\bar\rho$, i.e.
$\left|\frat{\partial\rho(x,z)}{\partial x}\right|\ll O(1)$) 
of initial instanton fields $G_{\mu\nu}$ are mainly taken into
consideration. Such a condition of smoothness or the adiabatic
change of instanton size dictates, in practice, 
another essential simplification to
define the field in the center of instanton   
$\frat{\partial\rho(x,z)}{\partial x}\sim
\left.\frat{\partial\rho(x,z)}{\partial x}\right|_{x=z}$ 
as a characteristic deformation field in further calculations.
 
In principle, the correction to the instanton field produced by the
quark presence might be calculated in its most general form,
if the gluon Green function in the instanton medium is known, as 
\begin{equation}
\label{gGF}
a^a_{\mu}(x,z)=\int d\xi~D^{ab}_{\mu\nu}(x-z,\xi-z)
~J^{b}_\nu(\xi-z_\psi)~,
\end{equation}
where $J^{b}_\nu$ is the current of external quark source, 
$z_\psi$ belongs to the region of long wavelength disturbance and, 
at last, $D^{ab}_{\mu\nu}$ is the Green function of PP in the instanton 
medium. 
In fact, this Green function (even when considered in the field of one
instanton) is not well defined \cite{car} but, seems, for the case in hands 
we could develop the selfconsistent way to calculate the regularized Green 
function. The nonsingular propagator behaviour in the soft momentum region is 
defined by the mass gap of phononlike excitations. Fortunately, the exact
form of the Green function occurs to be not of great importance here
(we are planning to return to the problem of regularized Green function 
calculation in the forthcoming publication). In the coordinate space it is
peaked around the average PP size being in nonperturbative regime
and, hence, integral Eq. (\ref{gGF}) is estimated to be
\begin{equation}
\label{est}
a^a_{\mu}(x,z)\simeq \bar D^{ab}_{\mu\nu}(x-z)~\bar J^b_{\nu}(z-z_\psi)~.
\end{equation}
On the other hand, the exact instanton definition in the singular gauge,
$$A^{a}_\mu(x,z)=-\frat{\bar\eta_{a\mu\nu}}{g}\frat{\partial}
{\partial x_\nu}\ln\left(1+\frat{\rho^2}{y^2}\right)~,$$
leads to the following correction to the instanton potential
\begin{equation}
\label{pop}
a^{a}_\mu(x)=H^{a}_{\mu\nu}(x,z)~\frat{\partial \rho}{\partial x_\nu}~,
\end{equation}
where $H^{a}_{\mu\nu}(x,z)=-\frat{\bar\eta_{a\mu\nu}}{g}
\frat{2\rho(x,z)}{y^2+\rho^2(x,z)}$.
Confining within the precision accepted here we could take 
$\rho(x,z)\simeq\bar\rho$, i.e.
$H^a_{\mu\nu}(x,z)=H^a_{\mu\nu}(x-z)$.

If we compare now both definitions of the correction we are capable to get
immediately for $\rho_\nu$ the following equation
$$H^a_{\mu\nu}(x-z)~\frat{\partial \rho(x,z)}{\partial x_\nu}=
\bar D^{ab}_{\mu\nu}(x-z)~\bar J^b_{\nu}(z-z_\psi)~.
$$
In fact, the current $\bar J$ might be taken constant in the long wavelength  
approximation. Then neglecting the current gradients one is allowed to
change the derivatives to obtain the following estimate of the deformation
velocity
$\left |\frat{\partial\rho(x,z)}{\partial x}\right |\sim
\left |\frat{\partial\rho(z)}{\partial z}\right |$
which looks to be well justified since there is no other fields in the
problem at all (in the adiabatic approximation). 
The contribution of deformed (anti-)instantons to the functional
integral
(when the corrections coming from the PP deformation fields are
absorbed) may be estimated as \cite{we}
$$
\langle S\rangle\simeq\int d z \int d\rho~n(\rho)
~\left\{~\frat{\kappa}{2}~\left(\frat{\partial \rho}{\partial z}\right)^2
+s(\rho)\right\}~,
$$
where $\kappa$ is the kinetic coefficient being derived within
the quasiclassical approach. Our estimate of it gives the value of a few 
single instanton actions $\kappa\sim c~\beta$ with the coefficient 
$c\sim 1.5\div 6$ depending quantitatively on the ansatz
supposed for the saturating configurations.  Although this estimate
is not much meaningful because there is no the vital 
$\kappa$ dependence eventually (becomes shortly clear) and the kinetic
term could be introduced phenomenologically. Thus, this coefficient 
should be fixed on a characteristic scale, for example 
$\kappa\sim\kappa(\bar\rho)$, if we not are planning to be beyond the 
precision peculiar to the approach. Actually, it means adding the 
small contribution of kinetic energy type to the action
per one instanton only. Such a term results from the scalar field
of deformations and affects negligibly the pre-exponential factors
of the functional integral. In one's turn the pre-exponential factors
do the negligible influence on the kinetic term as well. The deformation
fields induced by the dilations and rotations in the isotopic space 
result in the singular and trivial kinetic coefficients, respectively 
\cite{we}. Then to evaluate the mass scale related to those modes one
needs the heuristic ideas rooted, apparently, beyond the conventional
IL and SCSB.

If we strive for being within the approximation developed we should 
retain the small terms of the second order in deviation from the 
point of action minimum $\left.\frat{d s(\rho)}{d\rho}
\right|_{\rho=\rho_c}=0$ only expecting the approximate validity of 
\begin{equation}
\label{dec} 
s(\rho)\simeq s(\bar\rho)+\frat{s^{(2)}(\bar\rho)}{2}~\varphi^2~,
\end{equation} 
where 
$s^{(2)}(\bar\rho)\simeq\left.\frat{d^2 s(\rho)}{d\rho^2}
\right|_{\rho_c}=\frat{4\nu}{\overline{\rho^2}}$ 
and the scalar field
$\varphi=\delta\rho=\rho-\rho_c\simeq\rho-\bar\rho$
is the field of deviations from the equilibrium value of  
$\rho_c=\bar\rho~\left(1-\frat{1}{2\nu}\right)^{1/2}\simeq\bar\rho$. 
Consequently, the deformation field is described by the following 
Lagrangian density 
$$
{\cal L}=\frat{n\kappa}{2}
\left\{~\left(\frat{\partial \varphi}{\partial z}\right)^2+
M^2\varphi^2\right\}
$$
with the mass gap of the phononlike excitations
$$
M^2=\frat{s^{(2)}(\bar\rho)}{\kappa}=
\frat{4\nu}{\kappa \overline{\rho^2}}
$$
which is estimated for IL with $N_{c}=3$, for example, in the   
quenched approximation to be
$$M\approx 1.21~\Lambda~~$$   
if 
$c=4,~\bar\rho~\Lambda\approx 0.37,~\beta\approx 17.5,~n~\Lambda^{-4}
\approx 0.44$ 
(for the details see the tables of Appendix).  

Changing the variables we obtain the gluon part of the generating
functional as
$$
{\cal Z}^{'}_g\sim\int D\varphi~
\left|\frat{\delta A}{\delta \varphi\cdots}\right|~
\exp\left\{-\frat{n\kappa}{2}\int dz
\left[~\left( \frat{\partial \varphi}
{\partial z}\right)^2+M^2\varphi^2\right]\right\}~,
$$
with the Jacobian $\left|\frat{\delta A}{\delta \varphi\cdots}\right|$
corresponding those new variables introduced. 
Let us notice that the latter looks pretty formal because of incomplete
set of the transformations shown.
However, in the adiabatic approximation, as was mentioned above, 
the preexponentional Jacobian contribution to the action
{\footnote{Generally, the deformation field $\rho_\nu$ and
integration variable $a^a_{\mu}$ (\ref{pop}) are related via the rotation 
matrix: $\widetilde a^{a}_\mu=\Omega_{ab}\Phi^b_{\mu\nu}(x-z) \rho_\nu$
and in the long-length wave approximation $\Phi$ might be constant 
 $\Phi^b_{\mu\nu}(0)~$($x\sim z$). 
With the rotation matrix spanning the colour field $a_\mu=\Omega^{-1}
\widetilde a_\mu$
on the fixed axis we can conclude that the vectors 
$a^z_{\mu}$ and $\rho_\nu$ are, in fact, in one to one correspondence
(of course, being within one loop approximation and up to this unessential 
colour rotation). Thus, the Jacobian contribution turns out to be 
an unessential constant.}}
being a c-number should be omitted.

Analyzing the modifications which arise now in the quark determinant 
${\cal Z}_\psi$ we take into account the variation of fermion 
zero modes resulting from the instanton size perturbed 
$$\Phi_\pm(x-z,\rho)\simeq\Phi_\pm(x-z,\rho_c)+
\Phi_{\pm}^{(1)}(x-z,\rho_c)~\delta\rho(x,z)~,$$
where 
$\Phi^{(1)}_\pm(u,\rho_c)=\left.\frat{\partial\Phi_\pm(u,\rho)}
{\partial\rho}\right|_{\rho=\rho_c}$ 
and because of the adiabaticity  it is
valid $\delta\rho(x,z)\simeq$ $~\delta\rho(z,z)=\varphi(z)$. The
additional contributions of scalar field generate the corresponding
corrections in the factors of the kernels $Y^{\pm}$ of Eq. (\ref{6}) 
which are treated in the linear approximation in $\varphi$ and 
taking ~approximately $\rho_c=\bar\rho$, i.e.
\begin{eqnarray}
\label{8}
i\hat\partial_x \Phi_\pm(x-z,\rho)~
\Phi^\dagger_{\pm}(y-z,\rho)~i\hat\partial_y\simeq
\Gamma_\pm(x,y,z,\bar\rho)+\Gamma^{(1)}_\pm(x,y,z,\bar\rho)~\varphi(z)~,
\end{eqnarray}
here we introduced the notations
$$\Gamma_\pm(x,y,z,\bar\rho)=i\hat\partial_x \Phi_\pm(x-z,\bar\rho)
\Phi^\dagger_{\pm}(y-z,\bar\rho)(-i\hat\partial_y)~,$$
$$\Gamma^{(1)}_\pm(x,y,z,\bar\rho)=
i\hat\partial_x \Phi^{(1)}_\pm(x-z,\bar\rho)
\Phi^\dagger_{\pm}(y-z,\bar\rho)~(-i\hat\partial_y)+
i\hat\partial_x \Phi_\pm(x-z,\bar\rho)~\Phi^{\dagger (1)}_{\pm}(y-z,\bar\rho)~
(-i\hat\partial_y)$$
with $-i\hat\partial_y$ left acting operator
(the gradients of scalar field $\varphi$ are negligible according to 
the adiabaticity assumption again). It is a simple matter to verify that 
the right hand side of  Eq. (\ref{8}), being integrated over $dzdU$, 
generates the following kernel (in the momentum space)
\begin{equation}
\label{8a} 
\frat{1}{N_c}\left[(2\pi)^4~\delta(k-l)~\gamma_0(k,k)+
\gamma_1(k,l)~\varphi(k-l)\right]\nonumber
\end{equation}
with $\gamma_0(k,k)=G^2(k)~,~G(k)=2\pi\bar\rho F(k\bar\rho/2)~,
~\gamma_1(k,l)=G(k)G'(l)+G'(k)G(l)~,~\\
G'(k)=\left.\frat{d G(k)}{d\rho}\right|_{\rho=\bar\rho},~
F(x)=2x~[I_0(x)K_1(x)-I_1(x)K_0(x)]-2~I_1(x)K_1(x)$,
where $I_i,~K_i~(i=0,1)$ are the modified Bessel functions.

The functional integral of Eq. (\ref{6}) including the
phononlike component may be exponentiated in the momentum space 
{\footnote{In the metric space we have the nonlocal 
Lagrangian of the phononlike deformations  
$\varphi(z)$ interacting with the quark fields $\psi^\dagger,~\psi$, i.e.
\begin{eqnarray}
\label{ms}
{\cal L}&=&\int dx~\psi^\dagger(x)i\hat\partial_x \psi(x)-
\int dz~ \frat{n\kappa}{2}
 \left\{\left(\frat{\partial \varphi}{\partial z}\right)^2
+M^2 \varphi^2(z)\right\}+\nonumber\\
&+&
\frat{i\lambda_\pm}{N_c}\int dxdydz~dU~
\psi^\dagger(x)\{\Gamma_\pm(x,y,z,\bar\rho)+
\Gamma^{(1)}_\pm(x,y,z,\bar\rho)~\varphi(z)\}\psi(y)~.
\nonumber
\end{eqnarray}
The physical meaning of the basic phenomenon behind this Lagrangian seems
pretty transparent. The propagation of quark fields through the instanton 
medium is accompanied by the IL disturbance (the analogy with well known 
polaron problem embarrasses us strongly in this point).}} 
with the auxiliary integration over the 
$\lambda$-parameter (see, for example \cite{2}) 
\begin{eqnarray}
\label{9}
&&{\cal Z}_\psi\simeq\int \frat{d\lambda}{2\pi}~
\exp\left\{N\ln\left(\frat{N}{i\lambda VR}\right)-N\right\}\times\nonumber\\
&&\times\int D\psi^\dagger D\psi~ 
\exp\left\{\int \frat{dkdl}{(2\pi)^8}~\psi^\dagger(k)
\left[(2\pi)^4\delta(k-l)
\left(-\hat k+\frat{i\lambda}{N_c}\gamma_0(k,k)\right)+
\frat{i\lambda}{N_c}
\gamma_1(k,l)~\varphi(k-l)\right]~\psi(l)\right\}\nonumber
\end{eqnarray}
(we dropped out the factor normalizing to the free Lagrangian everywhere). 
It is pertinent to mention here the Diakonov-Petrov result comes to the play 
precisely if the scalar field is switched off.

In order to avoid a lot of the needless coefficients in the further 
formulae we introduce the dimensionless variables 
(momenta, masses and vertices) 
\begin{equation}
\label{sc}
\frat{k\bar\rho}{2}\to k~,~~~\frat{M\bar\rho}{2}\to M~,
~~\gamma_0\to\bar\rho^2\gamma_0~,~~\frat{1}{(n\bar\rho^4\kappa)^{1/2}}~
\gamma_1\to\bar\rho\gamma_1~,
\end{equation}
the fields in turn 
\begin{equation}
\label{sc1}
\varphi(k)\to (n\kappa)^{-1/2}\bar\rho^3\varphi(k)~,~~~~
\psi(k)\to \bar\rho^{5/2}\psi(k)~,
\end{equation}
and eventually for $\lambda$ we are using 
$\mu=\frat{\lambda\bar\rho^3}{2N_c}$. Then the generating functional 
takes the following form 
\begin{eqnarray}
\label{10}
&&{\cal Z}\simeq
\int d\mu~{\cal Z}^{''}_g\int D\psi^\dagger D\psi~D\varphi~
\exp\left\{n\bar\rho^4\left(\ln\frat{n\bar\rho^4}{\mu}-1\right)
-\int \frat{dk}{\pi^4}~\frat{1}{2}~\varphi(-k)~
4~[k^2+M^2]~\varphi(k)\right\}\times\nonumber\\
[-.2cm]
\\[-.25cm]
&&\times\exp\left\{\int \frat{dkdl}{\pi^8}~\psi^\dagger(k)
~2~\left[\pi^4\delta(k-l)(-\hat k+i\mu\gamma_0(k,k))+i\mu~
\gamma_1(k,l)~\varphi(k-l)\right]~\psi(l)\right\}~,\nonumber
\end{eqnarray}
where ${\cal Z}^{''}_g$ is a part of gluon component of the generating
functional which survives after expanding the action  per one instanton 
Eq. (\ref{dec}). The functional obtained describes the IL state influenced
by the quarks when all the terms containing the scalar field are collected
(see also Appendix). As mentioned above we believe this influence analogous 
to the back impact of phononlike deformations on the quark determinant
does not considerably change the numerical results of the IL and SCSB theory. 
Noninteracting part of phononlike excitation Lagrangian characterizes
the IL reaction on the external long-length wave perturbation
and, apparently, is its general feature independent
of perturbating field nature.

The presence of quark condensate makes hint for the  
appropriate scheme of approximate calculation of the generating
functional 
$$
\psi^\dagger\psi~\varphi=\langle\psi^\dagger\psi\rangle~\varphi+
\{\psi^\dagger\psi-\langle\psi^\dagger\psi\rangle\}~\varphi~.
$$

\section*{III. Tadpole approximation}
The formal integration over the scalar field leads us to 
the four fermion interaction and the functional integral
can not be calculated exactly. However, due to 
smallness of scalar field corrections we may find the effective 
Lagrangian substituting the condensate value in lieu of one of the pairs of 
quark lines (see Fig. 1a.)
$$\psi^\dagger (k)\psi(l)\to\langle\psi^\dagger (k)\psi(l)\rangle=
-\pi^4 \delta(k-l)~Tr~S(k)$$ 
where $S(k)$ is the quark Green function.
In such an approach the diagram with four fermion lines in the lowest order 
of the perturbation theory in $\mu$ is reduced to the two-legs diagram
with one 
tadpole contribution (there are two such contributions because of two 
possible ways of pairing) 
\begin{eqnarray}
\label{12}
&&~~2~(i\mu)^2
\int \frat{dkdl~dk'dl'}{\pi^{16}}~\gamma_1(k,l)~\gamma_1(k',l')
~~\psi^\dagger(k)\psi(l)~~\psi^\dagger(k')\psi(l')~
~\langle \varphi(k-l)~\varphi(k'-l')\rangle \simeq\nonumber\\
&&\simeq 4~\mu^2
\int \frat{dk}{\pi^4}~\gamma_1(k,k)~\psi^\dagger(k)\psi(k)
~\int \frat{dl}{\pi^4}~\gamma_1(l,l)~Tr~S(l)~D(0)~,\nonumber 
\end{eqnarray}
where the natural pairing definition was introduced
$$
\langle\varphi(k)~\varphi(l)\rangle=
\pi^4\delta(k+l)~D(k),~~D(k)=\frat{1}{4~(k^2+M^2)}~.
$$
It is obvious the factors surrounding $\psi^\dagger(k)\psi(k)$ have a 
meaning of an additional contribution to the dynamical mass
\begin{equation}
\label{kirmas}
m(k)=\mu~\gamma_1(k,k)
~(-2i\mu)
\int \frat{dl}{\pi^4}~\gamma_1(l,l)~Tr~S(l)~D(0)~,
\end{equation}
(the initial mass term contains the factor $2$ when the 
dimensionless variables are utilized, i.e. takes a form $2im$.) 

The contribution of the graph with all the quark lines paired (see Fig. 1b) 
$$
-2~\mu^2
\left[ \int \frat{dk}{\pi^4}~\gamma_1(k,k)~Tr~S(k)\right]^2
\pi^4\delta(0)D(0)=
-\frat{\mu^2}{2}~\frat{V}{\bar\rho^4}~\frat{\kappa}{\nu}
\left[ \int \frat{dk}{\pi^4}~\gamma_1(k,k)~Tr~S(k)\right]^{2}~,
$$
together with contribution of the graph (Fig. 1c) 
$$2~\mu^2\int \frat{dkdl}{\pi^8}~{\mbox{ Tr}}
 \gamma_1(k,l)~\gamma_1(l,k)~S(k)~S(l)~
D(k-l)~,$$
should be taken into account at the same order of the $\mu$ expansion
while calculating the saddle point equation.
Here we used the natural regularization of the 
$\delta$-function $\delta(0)=\frat{1}{\pi^4}
\frat{V}{\bar\rho^4}$ in the dimensionless units. Then the quark
determinant after integrating over the scalar field reads 
\begin{eqnarray}
\label{14}
&&{\cal Z}\sim \int d\mu\int D\psi^\dagger D\psi~
\exp\left\{n\bar\rho^4\left(\ln\frat{n\bar\rho^4}{\mu}-1\right)+
\frat{2N_c^{2}}{n\bar\rho^4\nu}~\frat{V}{\bar\rho^4}~
\mu^4~c^2(\mu)-\right.\nonumber\\
&&\left.-2N_c\mu^2\frat{V}{\bar\rho^4}
\int \frat{dkdl}{\pi^8}~\gamma^{2}_1(k,l)
\frat{(kl)-\Gamma(k)\Gamma(l)}{(k^2+\Gamma^2(k))(l^2+\Gamma^2(l))}
D(k-l)+
\int\frat{dk}{\pi^4}~\psi^\dagger(k)~2~
[-\hat k+i\Gamma(k)]~\psi(k)\right\}=\nonumber\\
&&=
\int d\mu~\exp\left\{n\bar\rho^4\left(\ln\frat{n\bar\rho^4}{\mu}-1\right)+
\frat{2N_c^{2}}{n\bar\rho^4\nu}~\frat{V}{\bar\rho^4}~\mu^4~c^2(\mu)-
\right.\nonumber\\[-.2cm]
\\[-.25cm]
&&-
\left.2N_c\mu^2\frat{V}{\bar\rho^4}
\int \frat{dkdl}{\pi^8}~\gamma^{2}_1(k,l)
\frat{(kl)-\Gamma(k)\Gamma(l)}{(k^2+\Gamma^2(k))(l^2+\Gamma^2(l))}
D(k-l)
+~\frat{V}{\bar\rho^4}~\int\frat{dk}{\pi^4}~{\mbox{ Tr}}~
\ln[-\hat k+i\Gamma(k)]\right\}~,\nonumber
\end{eqnarray}
where the vertex function is defined as
$$
\Gamma(k)=\mu~\gamma_0(k,k)+m(k)~,
$$
and we introduced the function $c(\mu)$ convenient for the practical
calculations 
$$
c(\mu)=-\frat{i(n\bar\rho^4\kappa)^{1/2}}{2\mu~N_c}~\int \frat{dk}{\pi^4}
~\gamma_1(k,k)~{\mbox{ Tr}}~S(k)~.
$$
As seen from Eq. (\ref{14}) the Green function of the quark field is
selfconsistently defined by the following equation  
$$
2~[-\hat k+i\Gamma(k)]~S(k)=-1~.
$$
Searching the solution in the form
$$
S(k)=~A(k)~\hat k+i~B(k)~,
$$
we get
$$A(k)=\frat12~\frat{1}{k^2+\Gamma^2(k)}~,~~~~~~
B(k)=\frat12~\frat{\Gamma(k)}{k^2+\Gamma^2(k)}~.
$$
Using Eq. (\ref{kirmas}) and the definitions of $\Gamma(k)$ and $B(k)$ 
we have the complete integral equation
$$
\Gamma(k)=\mu~\gamma_0(k,k)+N_c\frat{\kappa}{\nu}~\mu^2~\gamma_1(k,k)
~\int\frat{dl}{\pi^4}~\gamma_1(l,l)~\frat{\Gamma(l)}{l^2+\Gamma^2(l)}~,
$$
which drives to have the convenient representation of the solution
$$
\Gamma(k)=\mu~\gamma_0(k,k)+\frat{N_c}{(n\bar\rho^4 \kappa)^{1/2}}
\frat{\kappa}{\nu}~\mu^3~c(\mu)
~\gamma_1(k,k)~.
$$ 
What concerns the function $c(\mu)$ it is not a great deal to obtain 
$$
c(\mu)=\frat{(n\bar\rho^4\kappa)^{1/2}}{\mu}
~\int\frat{dk}{\pi^4}~\gamma_1(k,k)~\frat{\Gamma(k)}{k^2+\Gamma^2(k)}~,
$$ 
and, therefore, the complete integral equation for the function
$c(\mu)$ which is shown in Fig. 2 for $N_f=1$.  
Let us underline the $N_f$-dependence of the $c(\mu)$ function in the 
interval of $\mu$ determined by saddle point value is unessential. 
Then we easily obtain for the additional contribution to the 
dynamical quark mass
\begin{equation}
\label{mff}
m(k)=\frat{N_c}{(n\bar\rho^4 \kappa)^{1/2}}
\frat{\kappa}{\nu}~\mu^3~c(\mu)~\gamma_1(k,k)~,
\end{equation}
and see the cancellation of the kinetic coefficient $\kappa$ 
in $m$ if we remember the definition (\ref{sc}). 
Thus, it means the precise value of the coefficient is 
unessential as declared. 

We have the following equation for the saddle point of the 
functional of Eq. (\ref{14}) 
\begin{eqnarray}
\label{20}
&&\frat{n\bar\rho^4}{\mu}
-2N_c\int\frat{dk}{\pi^4}~\frat{[\Gamma^2(k)]'_{\mu}}{k^2+\Gamma^2(k)}
+2N_c\int \frat{dkdl}{\pi^8}~\left\{\frat{\mu^2\gamma^{2}_1(k,l)
[(kl)-\Gamma(k)\Gamma(l)]}{(k^2+\Gamma^2(k))(l^2+\Gamma^2(l))}
\right\}'_{\mu}D(k-l)-\nonumber\\
[-.2cm]
\\[-.25cm]
&&-\frat{2N^2_{c}}{n\bar\rho^4\nu}~[\mu^4~c^2(\mu)]'_{\mu}
-(n\bar\rho^4)^{'}_{\mu}\ln\frat{n\bar\rho^4}{\mu}=0~,\nonumber
\end{eqnarray} 
where the prime is attributed to the differentiation in $\mu$. 
Being interested in receiving a closed equation we need to know the 
derivative $c'(\mu)$ too. 
Then the definition of $c(\mu)$ above allows us to have
$$(1-\mu^2 A(\mu))~
c'(\mu)=2\mu~A(\mu)~c(\mu)+B(\mu)~,$$
where 
$$A(\mu)=\alpha(\mu)~\frat{N_c\kappa}{\nu}\int \frat{dk}{\pi^4}
\frat{\gamma^2_{1}(k)}
{k^2+\Gamma^2(k)}~\frat{k^2-\Gamma^2(k)}{k^2+\Gamma^2(k)}~,
$$
$$B=-\frat{2(n\bar\rho^4 \kappa)^{1/2}}{\mu^2}\int \frat{dk}{\pi^4}
\frat{\gamma_1(k)~ \Gamma^3(k)}
{(k^2+\Gamma^2(k))^2}~,
$$
and 
$$\alpha(\mu)=1-\mu^2 \frat{N_c}{\beta\xi^2}
\frat{\Gamma(\nu+1/2)}{\nu^{1/2}\Gamma(\nu)}~
\frat{c(\mu)}{n\bar\rho^4\left(n\bar\rho^4-\frat{\nu}{2\beta\xi^2}\right)}~.
$$
 Thereby the saddle point equation absorbs the effect
of the IL parameter modification ($n(\mu)$)
produced by equilibrium instanton size shift 
$\rho_c\sim\bar\rho$ 
which comes in the leading order from simple tadpole graph  
\begin{eqnarray}
\label{tdp}
&&2~i\mu
\int \frat{dkdl}{\pi^8}~\gamma_1(k,l)~(-\pi^4)~\delta(k-l)~Tr~S(k)
~\varphi(k-l)=\Delta\cdot~\varphi(0)~,\nonumber\\
[-.2cm]
\\[-.25cm]
&&\Delta=-2i\mu~
\int \frat{dk}{\pi^4}~\gamma_1(k,k)~Tr~S(k)=
\frat{4N_c}{(n\bar\rho^4\kappa)^{1/2}}~\mu^2 c(\mu)~,\nonumber
\end{eqnarray}
(remind here $\varphi=\rho-\rho_c$, 
and $\varphi(0)=\int dz~\varphi(z)$ is the scalar field in momentum
representation).

In the Table 1 the numerical results  (M.S.Z.) are shown for 
$N_f=1$ comparing to those Diakonov and Petrov (D.P.)
\begin{center}
Table 1.\\\vspace{0.3cm}
\begin{tabular}{|ccc|ccc|}
\hline
     &D.P.  &                                           &         
     &M.S.Z.&                                           \\\hline
$\mu$         &$M(0)$&$-i\langle\psi^\dagger\psi\rangle$&
$\mu$         &$M(0)$&$-i\langle\psi^\dagger\psi\rangle$\\\hline 
\dmn{5.27}{-3}&$359$ &$-(333)^3$                        &
\dmn{4.81}{-3}&$362$ &$-(322)^3$\\
\hline
\end{tabular}
\end{center}
The parameters indicated in the Table 1 are\\
the dynamical quark mass\\
$$
 M(0)= 2\Gamma(0)~\left(\frat{1}{\bar\rho}\right)~~~[MeV]
$$
and the quark condensate\\
$$-i\langle\psi^\dagger\psi\rangle=i~Tr~S(x)|_{x=0}=-2N_c~
\int \frat{dk}{\pi^4} \frat{\Gamma(k)}{k^2+\Gamma^2(k)}
~\left(\frat{1}{\bar\rho}\right)^3~~~[MeV]^3~.
$$
Through this paper the value of renormalization 
constant is fixed by $\Lambda=280~MeV$. 
Then the IL parameters are slightly different from their 
conventional values 
$\bar\rho\sim (600~MeV)^{-1},~\bar R\sim (200~MeV)^{-1}$ 
(see the corresponding tables in Appendix). 
However, with the minor $\Lambda$ variation the parameters could be
optimally fitted. 
As expected the change of quark condensate is insignificant, the
order of several $MeV$, what hints the existence of new soft energy scale  
established by the disturbance which accompanies the quark propagation through
the instanton medium.

\section*{IV. Multiflavour approach}
In order to match the approach developed to phenomenological estimates
we need the generalization for $N_f>1$.
Then the quark determinant becomes
\cite{2}, \cite{dp2} 
$$
{\cal Z}_\psi~\simeq~\int D\psi^\dagger D\psi~ 
\exp\left\{\int d x~\sum_{f=1}^{N_f}\psi_{f}^\dagger(x)~
i\hat\partial\psi_{f}(x)\right\}~
\left(\frat{Y^+}{VR^{N_f}}\right)^{N_+}~
\left(\frat{Y^-}{VR^{N_f}}\right)^{N_-}~,
$$
$$Y^{\pm}=i^{N_f} \int dz~dU~d\rho~n(\rho)/n
\prod_{f=1}^{N_f}\int dx_fdy_f~\psi_{f}^\dagger(x_f)~i\hat\partial_{x_f} 
\Phi_\pm(x_f-z)~
\Phi^\dagger_{\pm}(y_f-z)~i\hat\partial_{y_f}~\psi_{f}(y_f)~. 
$$
With phononlike component included every pair of the zero modes 
$\sim\Phi~\Phi^\dagger$ acquires the additional term similar to 
Eq. (\ref{8}). The appropriate transformation driving the factors $Y^\pm$ 
to their determinant forms \cite{2} is still valid here since the 
correction term differs from the basic one with the  scalar field 
$\varphi$.  
The complete integration over $dz$ leads (in the adiabatic approximation 
$\varphi(x,z)\to\varphi(z)$) to the transparent Lagrangian form with the 
momentum conservation of all interacting particles. 
Besides, we keep the main terms of $Y^\pm$ expanding in $\varphi$.
The quark zero modes generate the factor similar to Eq. (\ref{8a}) with  
$\frat{1}{N_c}$ being changed by the factor 
$\left(\frat{1}{N_c}\right)^{N_f}$ and then in the leading  $N_c$ order
we have 
\begin{eqnarray}
&&Y^\pm=\left(\frat{1}{N_c}\right)^{N_f}\int dz~
\det_{N_f}\left(i~J^{\pm}(z)\right)~,\nonumber\\
&&J^{\pm}_{fg}(z)=\int\frat{dkdl}{(2\pi)^8}
\left[e^{i(k-l)z}\gamma_0(k,l)+\int\frat{dp}{(2\pi)^4}~e^{i(k-l+p)z}
\gamma_1(k,l)~\varphi(p)\right]~
\psi^\dagger_{f}(k)~\frat{1\pm\gamma_5}{2}~\psi_g(l)~.
\nonumber
\end{eqnarray}
While providing the functional with the Gaussian form we perform
the integration over the auxiliary parameter $\lambda$ together with the 
bosonization resulting in the integration over the auxiliary matrix 
$N_f\times N_f$ meson fields \cite{dp2}
$$
\exp\left[\lambda~\det \left(\frat{i~J}{N_c}\right)\right]\simeq
\int d {\cal M}~\exp\left\{i~Tr[{\cal M}J]-(N_f-1)
\left(\frat{\det[{\cal M}N_c]}{\lambda}\right)^{\frac{1}{N_f-1}}
\right\}~.
$$
As a result the generating functional may be written as
\begin{eqnarray}
\label{24}
&&{\cal Z}=\int\frat{d\lambda}{2\pi}~{\cal Z}^{''}_g~
\exp(-N\ln\lambda)~\int D\varphi~
\exp\left\{-\int \frat{dk}{(2\pi)^4}~\frat{n\kappa}{2}~\varphi(-k)~
[k^2+M^2]~\varphi(k)\right\}\cdot\nonumber\\
&&\cdot \int D{\cal M}_{L,R}~\exp\int dz
\left\{-(N_f-1)\left[
\left(\frat{\det[{\cal M}_L N_c]}{\lambda}\right)^{\frac{1}{N_f-1}}+
\left(\frat{\det[{\cal M}_R N_c]}{\lambda}\right)^{\frac{1}{N_f-1}}
\right]\right\}\cdot\\
&&\cdot\int D\psi^\dagger D\psi~\exp 
\left\{\int \frat{dk}{(2\pi)^4} \sum_f\psi_{f}^\dagger(k)(-\hat k)\psi_f(k) 
+i~\int dz\left(Tr[{\cal M}_L J^+]+Tr[{\cal M}_R J^-]\right)\right\}~.
\nonumber
\end{eqnarray}
Now  scalar field interacts with the quarks of the different 
flavours, nevertheless,  the dominant contribution is expected  from 
the tadpole graphs where any pair of the quark fields is taken in the 
condensate approximation as done at $N_f=1$
$$
\psi^\dagger_{f}(k)\psi_g(l)\to\langle\psi^\dagger_{f}(k)\psi_g(l)\rangle=
-\pi^4\delta_{fg}~\delta(k-l)~Tr~S(k)~.
$$
As for the condensate itself we obtain it as the nontrivial solution
of saddle point equation. For example, it is
$$({\cal M}_{L,R})_{fg}={\cal M}~\delta_{fg}
$$
for the diagonal meson fields.
The dimensionless convenient variables (in addition to  
Eqs. (\ref{sc}), (\ref{sc1})) are the following
$$
\frat{{\cal M}}{2}~\bar\rho^3\to \mu~,
~~~~\left(\frat{\lambda\bar\rho^4}{(2N_c\bar\rho)^{N_f}}
\right)^{\frac{1}{N_f-1}}
\to g~.
$$
Then the effective action (${\cal Z}=\int dgd\mu~\exp \{-V_{eff}\}$) in new
designations has the form
\begin{eqnarray}
&&V_{\mbox{eff}}=N(N_f-1)\ln g-\frat{V}{\bar\rho^4}(N_f-1)
\frat{2\mu^{\frac{N_f}{N_f-1}}}{g}
-\frat{V}{\bar\rho^4}\frat{2N_f^{2}N_c^{2}}{n\bar\rho^4~\nu}\mu^4 c^2(\mu)+
\nonumber\\
[-.2cm]
\\[-.25cm]
&&+2N_cN_f\mu^2\frat{V}{\rho^4}
\int \frat{dkdl}{\pi^8}~\gamma^{2}_1(k,l)
\frat{(kl)-\Gamma(k)\Gamma(l)}{(k^2+\Gamma^2(k))(l^2+\Gamma^2(l))}
D(k-l)-
2N_fN_c~\frat{V}{\bar\rho^4}~\int \frat{dk}{\pi^4}\ln\{k^2+\Gamma^2(k)\}~,
\nonumber
\end{eqnarray}
and  the saddle point equation reads 
\begin{eqnarray}
\label{pereval}
&&\frat{n\bar\rho^4}{\mu}-2N_c
\int\frat{dk}{\pi^4}~\frat{[\Gamma^2(k)]'_{\mu}}{k^2+\Gamma^2(k)}-
\frat{2N_c^{2}N_f}{n\bar\rho^4\nu}[\mu^4~c^2(\mu)]'_{\mu}+\nonumber\\
[-.2cm]
\\[-.25cm]
&&+2N_c\int \frat{dkdl}{\pi^8}~\left\{\frat{\mu^2\gamma^{2}_1(k,l)
[(kl)-\Gamma(k)\Gamma(l)]}{(k^2+\Gamma^2(k))(l^2+\Gamma^2(l))}
\right\}'_{\mu}D(k-l)
-(n\bar\rho^4)^{'}_{\mu}
\ln\left(\left(\frat{n\bar\rho^4}{2}\right)^{1/N_f}
\frat{1}{\mu}\right)=0~.\nonumber
\end{eqnarray}
The additional contribution to the dynamical quark mass in Eq. (\ref{mff})
gains the factor $N_f$ because the scalar nature of the phononlike
field  requires to match the tadpole quark field condensates of all
$N_f$ flavours to each vertex
$$
m(k)=\frat{N_f N_c}{(n\bar\rho^4 \kappa)^{1/2}}
\frat{\kappa}{\nu}~\mu^3~c(\mu)~\gamma_1(k,k)~.
$$
Table 2 complements the Table 1 with the calculations at $N_f=2$
\begin{center}
Table 2.\\\vspace{0.3cm}
\begin{tabular}{|ccccc|ccccc|}
\hline
&D.P.&&&&M.S.Z.&&&&\\\hline
$\mu$&$M(0)$&$-i\langle\psi^\dagger\psi\rangle$&$F_{\pi}$&$F^{'}_\pi$
&$\mu$&$M(0)$&$-i\langle\psi^\dagger\psi\rangle$&$F_{\pi}$&$F^{'}_\pi$\\\hline
\dmn{4.83}{-3}&$386$&$-(381)^3$&$135$&$111$
&\dmn{3.26}{-3}&$298$&$-(335)^3$&$108$&$90$
\\\hline
\end{tabular}
\end{center}
where $F_{\pi}~~[MeV]$ is the pion decay constant and
$$
F_{\pi}^2=~\frat{N_c N_f}{2}~
\int \frat{dk}{\pi^4}~\frat{\Gamma^2(k)-\frat{k}{2}\Gamma'(k)\Gamma(k)
+\frat{k^2}{4}(\Gamma'(k))^2}{(k^2+\Gamma^2(k))^2}
~\left(\frat{1}{\bar\rho}\right)^2~,
$$
$F^{'}_\pi~~[MeV]$ is its approximated form 
$$
F^{'2}_\pi=\frat{N_c N_f}{2}~
\int \frat{dk}{\pi^4} \frat{\Gamma^2(k)}{(k^2+\Gamma^2(k))^2}
~\left(\frat{1}{\bar\rho}\right)^2~,
$$
here $\Gamma'(k)=\frat{d\Gamma(k)}{dk}$, 
and the condensate $-i\langle\psi^\dagger \psi\rangle$ is implied for the 
quarks of each flavour.

The light particle introduced and which imitates the scalar glueball
properties does not affect significantly the SCSB parameters and correctly
describes the soft pion excitations of quark condensate. Meanwhile, the 
experimental status of this light scalar glueball is very vague. We 
believe the phononlike excitations could manifest themselves being mixed,
for instance, with the excitations of the quark condensate in the scalar 
channel.

\section*{VI. Conclusion}
In this paper we have developed the consistent approach to describe the 
interaction of quarks with IL. Theoretically, it is based (and justified) 
on the particular choice of the configurations saturating the functional 
integral what is not merely a technical exercise. They are the 
deformable (crumpled) (anti-)instantons with the variable parameters 
$\gamma(x)$ and in the concrete treatment of this paper we play with the 
variation of the PP size $\rho(x,z)$. In a sense, such an ansatz is strongly 
motivated by the form of quark determinant which is solely dependent on the 
average instanton size in the SCSB theory. We have demonstrated  that in the 
long-length wave approximation the variational problem of the deformation 
field optimization turns into the construction of effective Lagrangian 
for the  scalar phononlike  $\varphi$ and quark fields with the Yukawa 
interaction. Physically, it allows us to analyze the backreaction of 
quarks on the instanton vacuum. We have pointed out this influence on the IL 
parameters as negligible. The modification of the SCSB parameters turns out
pretty poor as well. In particular, the scale of quark condensate change 
amounts to a few $MeV$ only. Nevertheless, switching on the phononlike 
excitations of IL leads to several qualitatively new and interesting effects. 
The propagation of the quark condensate disturbances over IL happens in this 
approach to be in close analogy with well-known polaron problem. We imply a 
necessity to take into account the medium feedback while elementary 
excitations  propagating.
Besides, it hints that fitting the parameters $\bar\rho,~n$ and 
renormalization constant $\Lambda$ all together with the alteration of 
$s(\rho)$ profile function we might achieve suitable agreement not only in 
the order of magnitude. The difficulties which confronted us here illuminate 
the fundamental problem of gluon field penetration into the vacuum 
(the instanton vacuum in this particular case) as the most principle one. 
Indeed, it is a real challenge to answer the  question about the strong 
interaction carrier in the soft momentum region. Perhaps, the light particle 
of scalar glueball properties which appears inherently in our approach and 
should manifest itself in the mixture with the excitation of quark condensate 
in scalar channel ($\sigma$-meson) is not bad candidate for that role. By the 
way, it could be experimentally observed
{\footnote{
We receive the effective Lagrangian of Yukawa type which admits the
estimate of quark-anti-quark bound state while adapted to the non-relativistic 
approximation and the inequality
$\frat{\mu \gamma_0(0)}{4\pi~M}\frat{\mu^2\gamma_1^{2}(0)}
{n\bar\rho^4~\kappa}\ge2$ if valid signals its appearance.
In the problem under consideration the left hand side of the inequality
is $O(1)$.}}
as a wide resonance.

Summarizing, we understand  our calculation can not pretend to the precise 
quantitative agreement with experimental data and see many things to be 
done. We are planning shortly to consider the problem of instanton profile 
\cite{sim}, to make more realistic description of the PP interaction and
to push our ansatz beyond the long-length wave approximation analyzing
more precisely 'instanton Jacobian' 
$\left|\frat{\delta A}{\delta \varphi}\right|$.

The authors have benefited from the discussions with many people
but especially with N.O. Agasyan, M.M. Musakhanov, Yu.A. Simonov.  
The paper was accomplished under the Grants 
CERN-INTAS 2000-349, NATO~2000-PST.CLG 977482. 
Two of us (S.V.M. and A.M.S.) acknowledge Prof. M. Namiki and the HUJUKAI 
Fund for the permanent financial support.

\section*{Appendix}
The contribution of the quark determinant to the IL action is given by
the tadpole diagram Eq. (\ref{tdp}) which takes the following form when 
returned to the dimensional variables (see, Eq. (\ref{sc1})) 
$$
\Delta \varphi=\Delta~\frat{(n\kappa)^{1/2}}{\bar\rho^3}~\varphi(0)=
\Delta~ (n\bar\rho^4\kappa)^{1/2}\int d\rho\frat{n(\rho)}{n}
~\int \frat{dz}{\bar\rho^4}~
\frat{\rho(z)-\rho_c}{\bar\rho}~.
$$
Then the IL action, Eq. (\ref{s}), acquires the additional term
$$
\langle S\rangle=
\int d z~n \left\{\langle s\rangle-
\langle \Delta'~\frat{\rho-\rho_c}{\bar\rho}\rangle\right\}~,
$$
where $\Delta'=\frat{4N_c}{n\bar\rho^4}~\mu^2 c(\mu)$
and the mean action per one instanton is given by the following
functional
$\langle s_1 \rangle= \int d\rho~s_1(\rho) n(\rho)/n$ 
with
$$s_1(\rho)=\beta(\rho)+5 \ln(\Lambda\rho)-\ln \widetilde \beta^{2N_c}+
\beta\xi^2\rho^2n\overline{\rho^2}-\Delta'~(\rho-\rho_c)/\bar\rho~.
$$ 
In order to evaluate the equilibrium parameters of IL we treat the 
maximum principle
$$\langle e^{-S}\rangle\ge\langle e^{-S_0}\rangle
e^{-\langle S-S_0\rangle}$$
adapting it to the simplest version (when the approximating functional
is trivial $S_0=0$). In a sense, this choice of the approximating 
functional should be a little 'worse' than in Ref. \cite{2}. Its only 
advantage comes from the possibility to get the explicit formulae for the 
IL parameters in lieu of solving the complicated transcendental equation.
In equilibrium the instanton size distribution function 
$n(\rho)$ should be dependent on the IL action only, i.e.
$n(\rho)=C e^{-s(\rho)}$ where $C$ is a certain constant. 
This argument corresponds to the maximum principle of Ref. \cite{2}. 
Indeed, if one is going to approach the functional (\ref{s}) as a local form 
$\langle s \rangle= \int d\rho~ s_1(\rho) n(\rho)/n$ 
where
$s_1(\rho)=\beta(\rho)+5 \ln(\Lambda\rho)-\ln \widetilde \beta^{2N_c}+
\beta\xi^2\rho^2n\overline{\rho^2}$,   
it makes the approach  selfconsistent. The functional difference 
$\langle s \rangle-\langle s_1 \rangle= \int d\rho~\{ s(\rho)- s_1(\rho)\}
e^{-s(\rho)}/n$ being varied over $s(\rho)$ 
leads then to the result 
$s(\rho)=s_1(\rho)+const$ keeping into the mind an arbitrary
normalization. The maximum principle results in getting the mean action per 
one instanton as the IL parameters function, for instance, average instanton 
size $\bar\rho$. The corrections generated by the 'shifting' terms turn out 
to be small and we consider them in the linear approximation in the deviation 
$\Delta$. The following schematic expansion exhibits how the major 
contribution appears
\begin{equation}
\label{a5}
\langle s_1\rangle=
\frat{\langle (s+\delta)~e^{-s-\delta}\rangle}
{\langle e^{-s-\delta}\rangle}
\simeq
\frat{\langle s~e^{-s}\rangle+
\langle \delta~e^{-s}\rangle}{\langle e^{-s}\rangle}+
\frat{\langle s~e^{-s}\rangle\langle \delta~e^{-s}\rangle-
\langle s~\delta~e^{-s}\rangle \langle e^{-s}\rangle}
{\langle e^{-s}\rangle^2}~,
\end{equation}
here $\delta(\Delta)$ stands for a certain small 'shifting' contribution and 
$s$ is the action generated by the gluon component only. The last term in 
Eq. (\ref{a5}) is small comparing to the first one and we ~ignore it. 
Then it is clear that evaluating the mean action per one instanton is 
permissible to hold the gluon condensate contribution $s$ only (without the
'shifting' term $\delta$) in the exponential. Hence, we have for the mean 
action per one instanton 
$\langle s_1 \rangle= \int d\rho~s_1(\rho) n_0(\rho)/n_0$,
and $n_0(\rho)$ is the distribution function which does not include the
'shifting' term
{\footnote{The 'shifting' term changes the mass of phononlike excitation 
insignificantly. The equilibrium instanton size as dictated by the
condition $\left.\frat{d s(\rho)}{d\rho}\right|_{\rho=\rho_c}=0$ is equal 
to  $\rho_c=(\alpha+\Delta'~\beta)~\bar\rho,
~\alpha=\left(1-\frat{1}{2\nu}\right)^{1/2},~
\beta=\frat{1}{4\nu}\left\{1-\alpha~\frat{\Gamma(\nu+1/2)}
{\nu^{1/2}\Gamma(\nu)}\right\}$ and the second derivative of action in 
the equilibrium point equals to
  $s^{''}(\rho_c)=\frat{4\nu}{\overline{\rho^2}}+
\frat{2\nu}{\overline{\rho^2}}~\Delta'\left\{\frat{\Gamma(\nu+1/2)}
{\nu^{3/2}\Gamma(\nu)}-\frat{1}{2\nu\alpha}\right\}$. 
Another source of corrections to the kinetic coefficient appears while one 
considers the instanton profile change $A\to A+a$
where the field of correction is 
$a\sim\left.\frat{\partial \rho(x,z)}{\partial x}\right |_{x=z}$. 
This mode could appear  within the superposition ansatz Eq. (\ref{2}) and 
leads to  the modification of quark zero mode ($D(A+a)\psi=0$). 
Fortunately, both corrections to the kinetic term are numerically small.}}.
It is possible to obtain for the average squared instanton size and
the IL density that
{\footnote{In order not to overload the formulae with the factors making the
results dimensionless, which are proportional to the powers of $\Lambda$, 
we drop them out hoping it does not lead to the misunderstandings.}}  
\begin{equation}
\label{a7}
r^2 \overline{\rho^2}=\nu~\left\{1+\frat{\Delta'}{r\bar\rho}~
\frat{\Gamma(\nu+1/2)}{2\nu\Gamma(\nu)}\right\}\simeq
\nu~\left\{1+\Delta'~\frat{\Gamma(\nu+1/2)}{2\nu^{3/2}\Gamma(\nu)}\right\}~,
\end{equation}
\begin{equation}
\label{a8}
n=C~C_{N_c}\widetilde \beta^{2N_c}\frat{\Gamma(\nu)}{2r^{2\nu}}~,
\end{equation}
where the parameter $r^2$ equals to
\begin{equation}
\label{a7a}
r^2=\beta \xi^2n\overline{\rho^2}~. 
\end{equation}
Expanding
$\ln\rho=
\ln\bar\rho+\frat{\rho-\bar\rho}{\bar\rho}+
\frat12\frat{(\rho-\bar\rho)^2}{\bar\rho^2}+\cdots$
and using Eq. (\ref{a7}) we can show that
$$\frat{\int d\rho~ n_0(\rho) \ln \rho}{\int d\rho~ n_0(\rho)}=\ln\bar\rho+
\Phi_1(\nu)~,~~
\frat{\int d\rho~ n_0(\rho) \rho}{\int d\rho~ n_0(\rho)}=\bar\rho+
\Phi_2(\nu)~,$$
where $\Phi_1,~\Phi_2$ are the certain function of $\nu$ independent of 
$\bar\rho$. Besides, the average squared instanton size within the precision
accepted obeys the equality
$r^2 \overline{\rho^2}=\Phi(\nu)$,
and $\Phi(\nu)$ is the function of $\nu$ only. Then the mean action per one
instanton looks like
$$\langle s_1 \rangle= -2N_c \ln \widetilde \beta+
(2\nu-1)~\ln\bar\rho+F(\nu)
$$
($F(\nu)$ is again the function of $\nu$ only and its explicit form is
unessential here). Calculating its maximum in $\bar\rho$ we receive
$$\bar\rho=\exp\left\{-\frac{2N_c}{2\nu-1}\right\}~,
~\beta=\frat{2bN_c}{2\nu-1}-\ln C_{N_c}~.
$$
Eqs. (\ref{a7}), (\ref{a7a}) allows us to receive the equation for packing
fraction parameter 
\begin{equation}
\label{a7aa}
(n\bar\rho^4)^2-\frat{\nu}{\beta\xi^2}~n\rho^4=
\frat{\Delta}{\beta \xi^2}~
\frat{\Gamma(\nu+1/2)}{2\sqrt{\nu}~\Gamma(\nu)}~.
\end{equation}
and for positive root we find 
$$n=\nu~\frat{e^{\frac{8N_c}{2\nu-1}}}{\beta\xi^2}
\left\{1+\Delta'~\frat{\Gamma(\nu+1/2)}{2\nu^{3/2}\Gamma(\nu)}
\right\}~,
$$
and handling Eq. (\ref{a8}) we determine the constant $C$.

The IL parameters come about close to the parameter values of the 
Diakonov-Petrov approach \cite{2} and are shown in the following
Table
\begin{center}
Table 3.\\\vspace{0.3cm}
\begin{tabular}{|cccc|cccc|}
\hline
&&D.P.&&&&~M.S.Z.~~&\\\hline
      $N_f$&$\bar\rho\Lambda$&$n/\Lambda^4$  &$\beta$
       &$N_f$&$\bar\rho\Lambda$&$n/\Lambda^4$  &$\beta$\\\hline
0 & 0.37  &0.44&17.48 
&0 & 0.37  &0.48&17.48\\      
1 & 0.30  &0.81&18.86 
&1 & 0.33  &0.70&18.11\\
2 & 0.24  &1.59&20.12  
&2 & 0.28  &1.13&18.91\\
\hline
\end{tabular}
\end{center}
Here $N_f$ is the number of flavours 
($N_f=0$ corresponds to the quenched approximation) and
$N_c=3$.
It is curious to notice the quark influence on the IL
equilibrium state  provokes the increase of the IL density. 
 
In Table 4 we demonstrate the mass gap magnitude $M$ and the wave length in
the 'temporal' direction $\lambda_4=M^{-1}$. To make it more indicative we
show also the average distance between PPs from which it is clear, indeed, 
that the adiabatic approximation $\lambda\ge L\sim \bar R>\bar\rho$ is 
valid for the long-length wave excitations of the $\pi$-meson type. 
All the parameters are taken at $\kappa=4\beta$ but the primed ones  
correspond to the kinetic term value of $\kappa=6\beta$. 
\begin{center}
Table 4.\\\vspace{0.3cm}
\begin{tabular}{|clclcl|}
\hline
$N_f$&$M\Lambda^{-1}$&$\lambda\Lambda$
&$M^{'}\Lambda^{-1}$&$\lambda'\Lambda$&$\bar R\Lambda$\\\hline
0&1.21&0.83&0.99&1.01&1.2\\      
1&1.34&0.75&1.09&0.91&1.1\\
2&1.45&0.69&1.18&0.84&0.97\\
\hline
\end{tabular}
\end{center}
here $\bar R=n^{-1/4}$ is the distance between PPs.

\newpage

\begin{tabular}{ll}
\begin{picture}(100,100)(10,20)
\put(20,0){\line(0,2){100}}
\put(20,50){\line(2,0){10}}
\put(35,50){\line(2,0){10}}
\put(50,50){\line(2,0){10}}
\put(65,50){\line(2,0){10}}
\put(50,0){ a)}
\put(95,50){\circle{100}}
\end{picture}
&
\hspace{3cm}
\begin{picture}(100,100)(10,20)
\put(0,50){\circle{100}}
\put(20,50){\line(2,0){10}}
\put(35,50){\line(2,0){10}}
\put(50,50){\line(2,0){10}}
\put(65,50){\line(2,0){10}}
\put(95,50){\circle{100}}
\put(50,0){ b)}
\end{picture}
\end{tabular}

\begin{tabular}{ll}
\begin{picture}(100,100)(10,20)
\put(75,50){\line(2,0){10}}
\put(90,50){\line(2,0){10}}
\put(105,50){\line(2,0){10}}
\put(50,0){ c)}
\put(95,50){\circle{200}}
\end{picture}
&
\begin{picture}(100,100)(10,20)
\put(70,30){\line(2,0){10}}
\put(85,30){\line(2,0){10}}
\put(100,30){\line(2,0){10}}
\put(115,30){\line(2,0){10}}
\put(70,60){\line(2,0){55}}
\put(140,30){ phononlike field $\varphi$}
\put(140,60){ fermion fields $\psi,~\psi^\dagger$}
\end{picture}
\end{tabular}

\vspace{1.cm}

\begin{figure}[h]
\caption{The tadpole graphs.}
\vspace{1.cm}
\end{figure}

\begin{figure}[h]
\grpicture{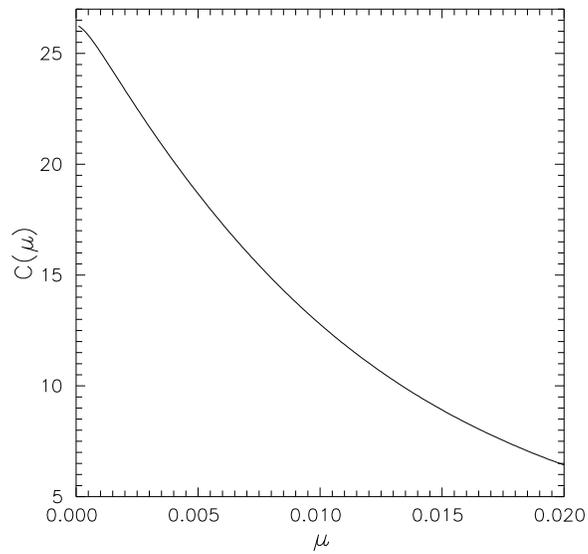}
\caption{The function $c(\mu)$ at $N_f=1$.}
\end{figure}

\end{document}